\documentclass[twocolumn,english,superscriptaddress]{revtex4-1}
\usepackage[T1]{fontenc}
\usepackage[latin9]{inputenc}
\usepackage{amsbsy}
\usepackage{amstext}
\usepackage{graphicx}

\makeatletter
\usepackage{babel}

\makeatother

\usepackage{babel}
\begin{document}
\title{Momentum dependence of $\phi$ meson's spin alignment}
\author{Xin-Li Sheng}
\affiliation{INFN-Firenze, Via Giovanni Sansone, 1, 50019 Sesto Fiorentino FI,
Italy}
\author{Shi Pu}
\affiliation{Department of Modern Physics, University of Science and Technology
of China, Hefei, Anhui 230026, China}
\author{Qun Wang}
\affiliation{Department of Modern Physics, University of Science and Technology
of China, Hefei, Anhui 230026, China}
\affiliation{School of Mechanics and Physics, Anhui University of Science and Technology,
Huainan,Anhui 232001, China}
\begin{abstract}
We study the rapidity and azimuthal angle dependences of the global
spin alignment $\rho_{00}$ for $\phi$ mesons with respect to the
reaction plane in Au+Au collisions at RHIC by the relativistic coalescence
model in the spin transport theory. The global spin alignment of $\phi$
mesons arises from local fluctuations of strong force fields whose
values are extracted from the STAR's data. The calculated results
show that $\rho_{00}<1/3$ at the rapidity $Y=0$, and then it increases
with rapidity and becomes $\rho_{00}>1/3$ at $Y=1$. Such a rapidity
dependence is dominated by the relative motion of the $\phi$ meson
in the bulk matter. We also give prediction for the azimuthal angle
dependence of $\rho_{00}$ at different rapidities.
\end{abstract}
\maketitle

\section{Introduction}

In noncentral heavy-ion collisions, the colliding nuclei carry a global
orbital angluar momentum (OAM). A small portion of the global OAM
is transferred into the quark-gluon plasma (QGP) in the form of vorticity
fields. Quarks are then polarized by vorticity fields through spin-orbit
couplings and form polarized hadrons by coalescence or recombination
\citep{Liang:2004ph,Voloshin:2004ha,Betz:2007kg,Becattini:2007sr,Gao:2007bc},
see Refs. \citep{Wang:2017jpl,Florkowski:2018fap,Gao:2020lxh,Huang:2020dtn,Becattini:2020ngo,Becattini:2022zvf}
for recent reviews. The effect of the vorticity field is supported
by the global spin polarization of $\Lambda$ and $\overline{\Lambda}$
hyperons observed by the STAR collaboration in Au+Au collisions \citep{STAR:2017ckg,STAR:2018gyt}.
According to the spin-flavor wave functions of $\Lambda$ and $\overline{\Lambda}$
in the quark model \citep{Liang:2004ph,Yang:2017sdk}, the spins of
$\Lambda$ and $\overline{\Lambda}$ are carried by $s$ and $\overline{s}$
quarks respectively, indicating that strange quarks in the QGP are
also globally polarized.


It was proposed by Liang and Wang in 2005 that the polarized quarks
can also form vector mesons by recombination with nonvanishing spin
alignment \citep{Liang:2004xn}. The spin of a vector meson is described
by a $3\times3$ spin density matrix $\rho$ with elements $\rho_{\lambda_{1}\lambda_{2}}$,
where subscripts $\lambda_{1},\lambda_{2}=0,\pm1$ label the spin
quantum number along a specific spin quantization direction. The spin
alignment refers to the element $\rho_{00}$, which denotes the probablity
for the spin state with $\lambda=0$. The polarization vector of the
vector meson is preferrably aligned in the spin quantization direction
when $\rho_{00}>1/3$, while it is preferrably aligned in the perpendicular
direction when $\rho_{00}<1/3$. In experiments, the spin alignment
can be measured through polar angle distributions of daughter particles
in p-wave strong decays, such as $\phi\rightarrow K^{+}+K^{-}$ \citep{Schilling:1969um,Liang:2004xn,Yang:2017sdk,ALICE:2019aid,STAR:2022fan},
or dilepton decays, such as $J/\psi\rightarrow\mu^{+}+\mu^{-}$ \citep{Faccioli:2010kd,DeMoura:2023jzz,ALICE:2020iev,ALICE:2022dyy}.


Rencently, the STAR collaboration has measured the global spin alignment
of $\phi$ mesons with respect to the reaction plane in Au+Au collisions
\citep{STAR:2022fan}, and the results show a significant positive
deviation from 1/3. Such a deviation is much larger than contributions
from conventional mechanisms such as vorticity fields and magnetic
fields \citep{Liang:2004xn,Yang:2017sdk,Xia:2020tyd,Sheng:2022ssp,Wei:2023pdf}.
Other possible contributions are proposed in \citep{Gao:2021rom,Muller:2021hpe,Li:2022vmb,Wagner:2022gza,Kumar:2023ghs}
but without quantitative results that can be compared with experimental
data. Upto now, the effect of local correlation or fluctuation of
a kind of the strong force field called the $\phi$ field \citep{Sheng:2019kmk,Sheng:2022ffb,Sheng:2022wsy}
is the only mechanism that can quantitatively explain the SATR data.
According to the chiral quark model \citep{Manohar:1983md,Fernandez:1993hx,Li:1997gd,Zhao:1998fn,Zacchi:2015lwa,Zacchi:2016tjw},
the SU(3) octet vector fields in the form of a $3\times3$ matrix
can be induced by vector currents of pseudo-Goldstone bosons that
surround $s$ and $\overline{s}$ quarks in the hadronization stage.
The $\phi$ field is just the ``33'' component of the SU(3) octet
vector fields that is coupled to $s$ and $\overline{s}$.


In this paper, we study the momentum dependence of $\rho_{00}^{y}$
in the relativistic coalescence model based on the spin Boltzmann
equation in non-equilibrium transport theory \citep{Sheng:2022wsy,Sheng:2022ffb}.
Here the superscript $i=x,y,z$ in $\rho_{00}^{i}$ denotes the spin
quantization direction. In this paper, we choose the beam direction
along the $z$-axis, the normal direction of the reaction plane along
the $y$-axis, and the impact parameter along the $x$-axis. The parameters
for the $\phi$ field's local fluctuations are extracted by fitting
the STAR data for $\rho_{00}^{y}$ and $\rho_{00}^{x}$ as functions
of collision energy \citep{STAR:2022fan}, and the values of these
parameters are the same as in Ref. \citep{Sheng:2022wsy}. Some recent
reviews on the STAR experiment and theoretical models can be found
in Refs. \citep{Chen:2023hnb,Wang:2023fvy,Sheng:2023act-aphys-sin}.


The paper is organized as follows. In Sec. \ref{sec:Theoretical-setup}
we give the formula for $\rho_{00}$ as functions of the $\phi$ field's
fluctuations. Assuming that the fluctuations are isotropic in the
lab frame, we derive a compact formula for the momentum dependence
of $\rho_{00}$. In Sec. \ref{sec:Numerical-results} we compute the
rapidity and azimuthal angle dependence of the global spin alignment
$\rho_{00}^{y}$ for $\phi$ mesons in Au+Au collisions. Finally we
make a summary and a discussion on the our result in Sec. \ref{sec:Summary}.
As notational convention, we use boldface symbols for three-dimensional
vectors, such as ${\bf p}=(p_{x},p_{y},p_{z})$.


\section{Theoretical model}

\label{sec:Theoretical-setup}In non-relativistic quark coalescence
model, the spin alignment of the $\phi$ meson depends on the spin
polarizations of its constituent quark and antiquark \citep{Liang:2004xn,Yang:2017sdk,Sheng:2019kmk}.
In Refs. \citep{Sheng:2022wsy,Sheng:2022ffb}, some of us constructed
a relativistic quark coalescence model based on spin Boltzmann equation
in transport theory to describe the spin phenomena of vector mesons.
Considering the quark polarization induced by $F_{\phi}^{\mu\nu}$,
the field strength tensor of the $\phi$ field, the spin alignment
for the $\phi$ meson is given by \citep{Sheng:2022wsy,Sheng:2022ffb}
\begin{eqnarray}
\rho_{00}(x,{\bf p}) & = & \frac{1}{3}-\frac{4g_{\phi}^{2}}{m_{\phi}^{2}T_{\text{h}}^{2}}C_{1}\left[\frac{1}{3}{\bf B}_{\phi}^{\prime}\cdot{\bf B}_{\phi}^{\prime}-(\boldsymbol{\epsilon}_{0}\cdot{\bf B}_{\phi}^{\prime})^{2}\right]\nonumber \\
 &  & -\frac{4g_{\phi}^{2}}{m_{\phi}^{2}T_{\text{h}}^{2}}C_{1}\left[\frac{1}{3}{\bf E}_{\phi}^{\prime}\cdot{\bf E}_{\phi}^{\prime}-(\boldsymbol{\epsilon}_{0}\cdot{\bf E}_{\phi}^{\prime})^{2}\right]\,,\label{eq:spin_alignment}
\end{eqnarray}
where $C_{1}$ and $C_{2}$ are two coefficients depending on $m_{\phi}$
(the $\phi$ meson's mass) and $m_{s}$ (mass of constituent strange
quark), $T_{\text{h}}$ is the local temperature at the hadronization
time, and $\boldsymbol{\epsilon}_{0}$ denotes the unit vector along
the measuring (spin quantization) direction which is also the $\phi$
meson's polarization vector for the spin state $\lambda=0$. Here
${\bf E}_{\phi}^{\prime}$ and ${\bf B}_{\phi}^{\prime}$ are electric
and magnetic parts of the $\phi$ field in the meson's rest frame,
which are functions of spacetime. When boosted to the lab frame, ${\bf E}_{\phi}^{\prime}$
and ${\bf B}_{\phi}^{\prime}$ also depend on the meson's momentum
${\bf p}$, i.e., they can be expressed in terms of fields in the
lab frame ${\bf B}_{\phi}$ and ${\bf E}_{\phi}$ as 
\begin{eqnarray}
{\bf B}_{\phi}^{\prime} & = & \gamma{\bf B}_{\phi}-\gamma{\bf v}\times{\bf E}_{\phi}+(1-\gamma)\frac{{\bf v}\cdot{\bf B}_{\phi}}{v^{2}}{\bf v}\,,\nonumber \\
{\bf E}_{\phi}^{\prime} & = & \gamma{\bf E}_{\phi}+\gamma{\bf v}\times{\bf B}_{\phi}+(1-\gamma)\frac{{\bf v}\cdot{\bf E}_{\phi}}{v^{2}}{\bf v}\,,\label{eq:Lorentz_transform}
\end{eqnarray}
where $\gamma=E_{{\bf p}}^{\phi}/m_{\phi}$ is the Lorentz contraction
factor and ${\bf v}={\bf p}/m_{\phi}$ is the $\phi$ meson's velocity.
Substituting Eq. (\ref{eq:Lorentz_transform}) into Eq. (\ref{eq:spin_alignment}),
we are able to express $\rho_{00}$ in terms of ${\bf B}_{\phi}$
and ${\bf E}_{\phi}$ which depend only on spacetime but not on momentum.


We learn from Eq. (\ref{eq:spin_alignment}) that the deviation from
$1/3$ for $\rho_{00}$ is caused by the anisotropy of local field
fluctuations in the meson's rest frame. The deviation is positive
(negative) when the fluctuation of ${\bf B}_{\phi}^{\prime}$ or ${\bf E}_{\phi}^{\prime}$
in the measuring direction $\boldsymbol{\epsilon}_{0}$ is larger
(smaller) than the average fluctuation in directions perpendicular
to $\boldsymbol{\epsilon}_{0}$. We assume that the fluctuations in
the lab frame are parameterized in an anisotropic form, 
\begin{equation}
\left\langle g_{\phi}^{2}{\bf B}_{\phi}^{i}{\bf B}_{\phi}^{j}/T_{\text{h}}^{2}\right\rangle =\left\langle g_{\phi}^{2}{\bf E}_{\phi}^{i}{\bf E}_{\phi}^{j}/T_{\text{h}}^{2}\right\rangle =F^{2}\delta^{ij}+\Delta\hat{{\bf a}}^{i}\hat{{\bf a}}^{j}\,,\label{eq:fluctuations_lab_frame}
\end{equation}
while the correlation between ${\bf B}_{\phi}$ and ${\bf E}_{\phi}$
is neglected, $\left\langle g_{\phi}^{2}{\bf B}_{\phi}^{i}{\bf E}_{\phi}^{j}/T_{\text{h}}^{2}\right\rangle =0$.
Here $\left\langle \cdots\right\rangle $ denotes the spacetime average,
$\hat{{\bf a}}$ denotes the direction of anisotropy, $F^{2}$ is
the isotropic part of the fluctuation, and $\Delta$ denotes the difference
between the fluctuation in $\hat{{\bf a}}$ and the average fluctuation
in directions perpendicular to $\hat{{\bf a}}$. The field configuration
in Eq. (\ref{eq:fluctuations_lab_frame}) becomes isotropic when $\Delta=0$.
The field fluctuations in the meson's rest frame can be obtained using
Eqs. (\ref{eq:Lorentz_transform}) and (\ref{eq:fluctuations_lab_frame}).


By substituting Eq. (\ref{eq:Lorentz_transform}) into Eq. (\ref{eq:spin_alignment}),
averaging over spacetime, and then applying Eq. (\ref{eq:fluctuations_lab_frame})
with $\Delta=0$, we obtain the spin alignment in the $y$-direction,
the direction of the global OAM with $\boldsymbol{\epsilon}_{0}=(0,1,0)$,
\begin{equation}
\left\langle \delta\rho_{00}^{y}\right\rangle ({\bf p})=\frac{8}{3m_{\phi}^{4}}(C_{1}+C_{2})F^{2}\left(\frac{p_{x}^{2}+p_{z}^{2}}{2}-p_{y}^{2}\right)\,,\label{eq:deviation_spin_alignemnt}
\end{equation}
where $\delta\rho_{00}^{y}\equiv\rho_{00}^{y}-1/3$ and $(C_{1}+C_{2})F^{2}$
is a positive number. Obviously, for static $\phi$ mesons with ${\bf p}=0$
we should have $\rho_{00}^{y}=1/3$ because field fluctuations are
isotropic in the lab frame. However, the motion of the $\phi$ meson
will break the symmetry. According to the Lorentz transformation of
fields in Eq. (\ref{eq:Lorentz_transform}), field components perpendicular
to the motion direction are enhanced by the $\gamma$ factor, while
the component in the motion direction is not. Therefore when observing
in the meson's rest frame, the fluctuation in the direction of motion
will be smaller than fluctuations in perpendicular directions. For
a meson with $p_{x}=p_{z}=0$ and $p_{y}\neq0$, we obtain the relation
$\left\langle ({\bf B}_{\phi,y}^{\prime})^{2}\right\rangle <\left\langle ({\bf B}_{\phi,x}^{\prime})^{2}\right\rangle =\left\langle ({\bf B}_{\phi,z}^{\prime})^{2}\right\rangle $,
leading to $\rho_{00}^{y}<1/3$ according to Eq. (\ref{eq:spin_alignment})
or (\ref{eq:deviation_spin_alignemnt}). Similarly, motions along
$x$- and $z$-directions lead to $\rho_{00}^{y}>1/3$.


We can also rewrite the result in Eq. (\ref{eq:deviation_spin_alignemnt})
in terms of the transverse momentum $p_{T}$, the azimuthal angle
$\varphi$, and the rapidity $Y$, 
\begin{equation}
\left\langle \delta\rho_{00}^{y}\right\rangle ({\bf p})\propto\frac{1}{2}p_{T}^{2}\left[3\cos(2\varphi)-1\right]+\sqrt{m_{\phi}^{2}+p_{T}^{2}}\sinh^{2}Y\,.\label{eq:deviation_pt_varphi_y}
\end{equation}
The azimuthal angle dependence shows a $\cos(2\varphi)$ structure,
which was first derived in Ref. \citep{Sheng:2022wsy}. We also find
that the spin alignment increases with rapidity. This is because the
anisotropy of field fluctuations in the meson's rest frame becomes
more significant in the direction of the meson's larger momentum.


We note that Eqs. (\ref{eq:deviation_spin_alignemnt}) and (\ref{eq:deviation_pt_varphi_y})
are based on the assumption (\ref{eq:fluctuations_lab_frame}) with
vanishing anisotropy $\Delta=0$ in the lab frame. A nonzero $\Delta$
will contribute to $\rho_{00}^{y}$, but the analytical formula is
too complicated to be given here. Our numerical results for nonzero
$\Delta$ in the next section will show that its contribution is small
compared with the results for $\Delta=0$. So the spin alignment is
dominated by the isotropic part of the field fluctuations.


\section{Numerical results}

\label{sec:Numerical-results}Similar to the previous work \citep{Sheng:2022wsy}
by some of us, we consider the anisotropy with respect to the $z$-direction
or the beam direction in heavy-ion collisions. So we assume that transverse
and longitudinal fluctuations are different $F_{T}^{2}=F^{2}$ and
$F_{z}^{2}=F^{2}+\Delta$, which are regarded as parameters that can
be extracted from the STAR's data on momentum integrated $\rho_{00}^{y}$
\citep{STAR:2022fan}. At collision energies $\sqrt{s_{\text{NN}}}=$
11.5, 19.6, 27, 39, 62.4, and 200 GeV, the extracted values of $F^{2}$
and $\Delta$ are $F^{2}/m_{\pi}^{2}=$ 16.5, 3.74, 3.42, 1.02, 2.85,
0.359 and $\Delta/m_{\pi}^{2}=$ 3.33, $-0.468$, $-0.9$, 0.218,
0.336, 0.128, respectively.

\begin{figure}
\includegraphics[width=8cm]{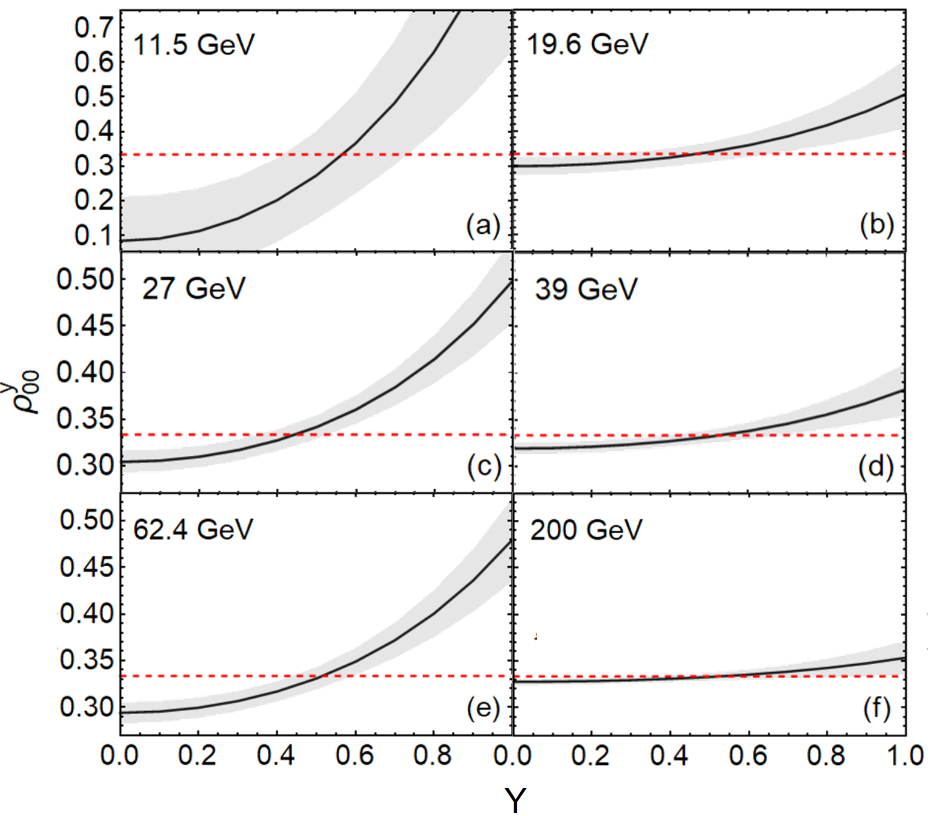}

\caption{\label{fig:spin_alignment_rapidity}The global spin alignment $\rho_{00}^{y}$
(solid lines) as functions of rapidity at collision energies $\sqrt{s_{\text{NN}}}=$11.5,
19.6, 27, 39, 62.4, and 200 GeV. The shaded areas are error bands
from fitting parameters for local fluctuations in strong force fields.
The dashed lines indicate the value of 1/3 without spin alignment.}
\end{figure}

\begin{figure}
\includegraphics[width=8cm]{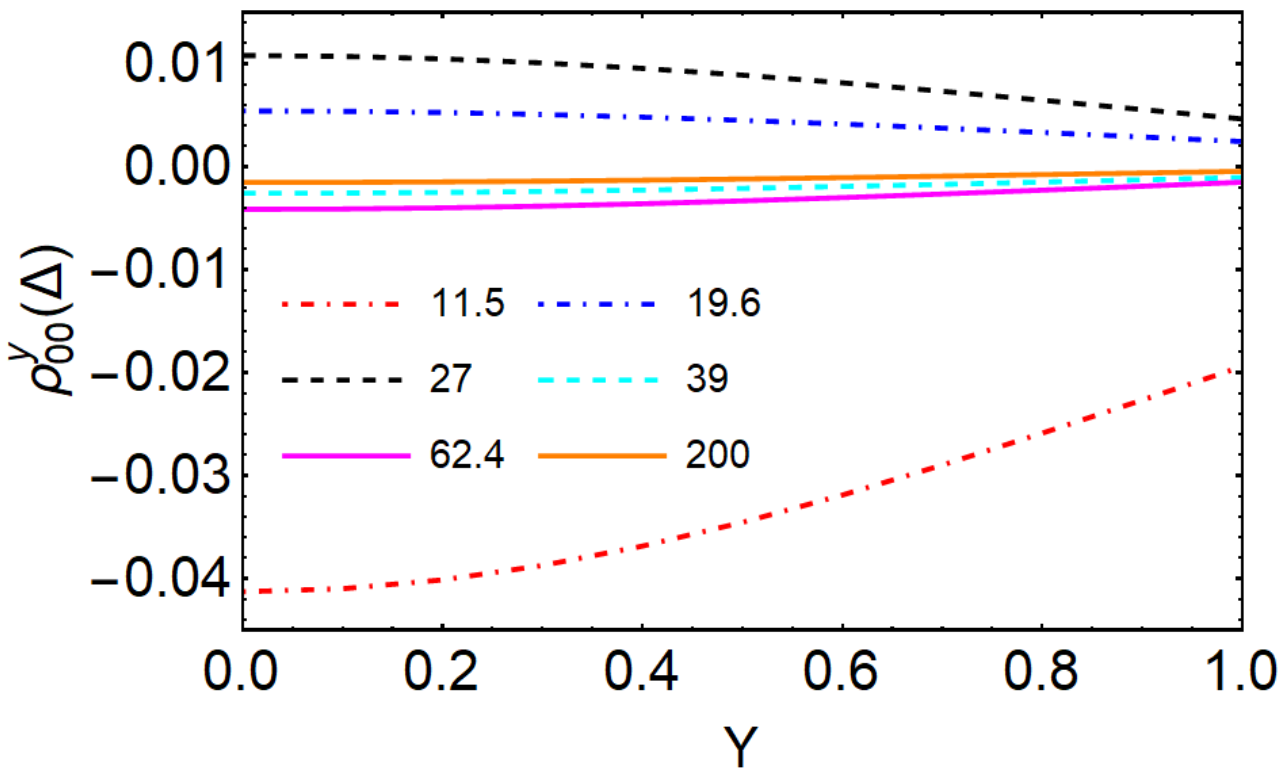}

\caption{\label{fig:contribution_delta}Contributions from the anisotropy parameter
$\Delta$ in the lab frame to the $\rho_{00}^{y}$ versus rapidity
at different collision energies.}
\end{figure}

In Fig. \ref{fig:spin_alignment_rapidity}, we present $\rho_{00}^{y}$
as functions of the $\phi$ meson's rapidity $Y$ at collision energies
$\sqrt{s_{\text{NN}}}=$11.5 - 200 GeV. At each rapidity and energy,
we assume that the spacetime averages of fluctuations follow Eq. (\ref{eq:fluctuations_lab_frame}).
We take averages over the meson's transverse momentum in the range
1.2 GeV$<p_{T}<$5.4 GeV and the azimuthal angle in the range $0<\varphi<2\pi$,
weighted by the $\phi$ meson's momentum spectra 
\begin{equation}
E_{{\bf p}}\frac{d^{3}N}{d^{3}{\bf p}}=\frac{d^{2}N}{2\pi p_{T}dp_{T}dY}\left[1+2v_{2}(p_{T})\cos(2\varphi)\right],
\end{equation}
where the transverse momentum spectra and $v_{2}(p_{T})$ are taken
from STAR's data \citep{STAR:2007mum,STAR:2008bgi,STAR:2013ayu,STAR:2019bjj}.
At all energies, the derivation from 1/3 is negative at the mid-rapidity
$Y=0$, while it increases at larger $Y$. To see the influence of
the anisotropy in the field fluctuation, we isolate the contribution
from $\Delta$ to $\rho_{00}^{y}$ by taking the difference between
results in Fig. \ref{fig:spin_alignment_rapidity} with non-vanishing
$\Delta$ and those obtained with the same $F^{2}$ but $\Delta=0$.
As shown in Fig. \ref{fig:contribution_delta}, the effects of non-vanishing
$\Delta$ are one order of magnitude smaller than $\rho_{00}^{y}-1/3$
in Fig. \ref{fig:spin_alignment_rapidity}. This indicates that the
rapidity dependence in Fig. \ref{fig:spin_alignment_rapidity} is
dominated by the isotropic part $F^{2}$ in Eq. (\ref{eq:fluctuations_lab_frame}).


The theoretical model in Sec. \ref{sec:Theoretical-setup} provides
a clear picture for the rapidity dependence of the spin alignment.
The bulk matter in which $\phi$ mesons are produced can be treated
as a nearly-isotropic medium with a small anisotropy along the $z$-direction
in fluctuations of strong force fields described by Eq. (\ref{eq:fluctuations_lab_frame}),
i.e. $F^{2}\gg\Delta$. The $\phi$ meson's motion relative to the
bulk matter breaks the rotational symmetry in meson's rest frame,
leading to a larger probability for spin $\pm1$ states than the spin
0 state with respect to the motion direction. For example, if mesons
move in the $z$-direction, we have $\rho_{00}^{z}<1/3$, or equivalently
$\rho_{00}^{x}=\rho_{00}^{y}>1/3$ because of the normalization condition
$\rho_{00}^{x}+\rho_{00}^{y}+\rho_{00}^{z}=1$.

\begin{figure}
\includegraphics[width=8cm]{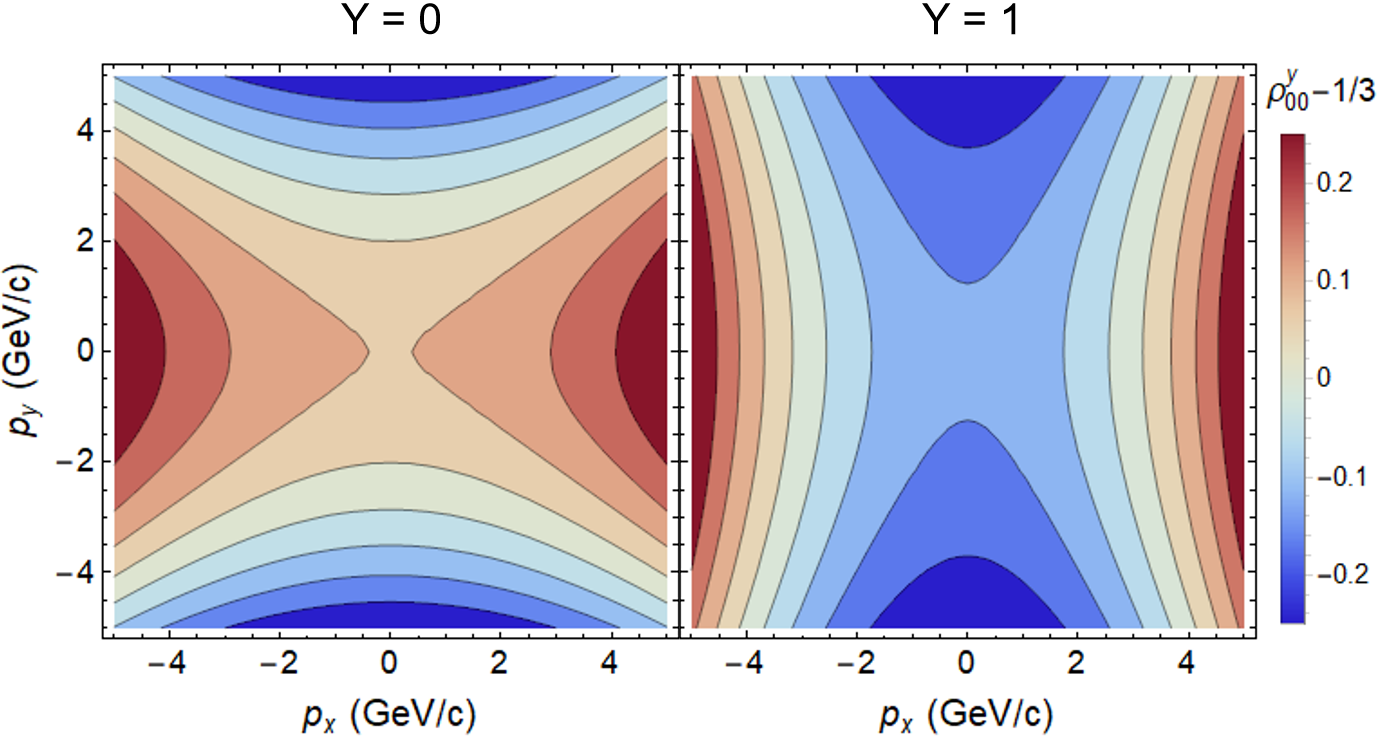}

\caption{\label{fig:rho00_px_py}The contour plots for $\delta\rho_{00}^{y}$
in the transverse momentum plane $(p_{x},p_{y})$ for $\phi$ mesons
at $Y=0$ (left panel) and $Y=1$ (right panel) in Au+Au collisions
at 200 GeV. }
\end{figure}

\begin{figure}
\includegraphics[width=8cm]{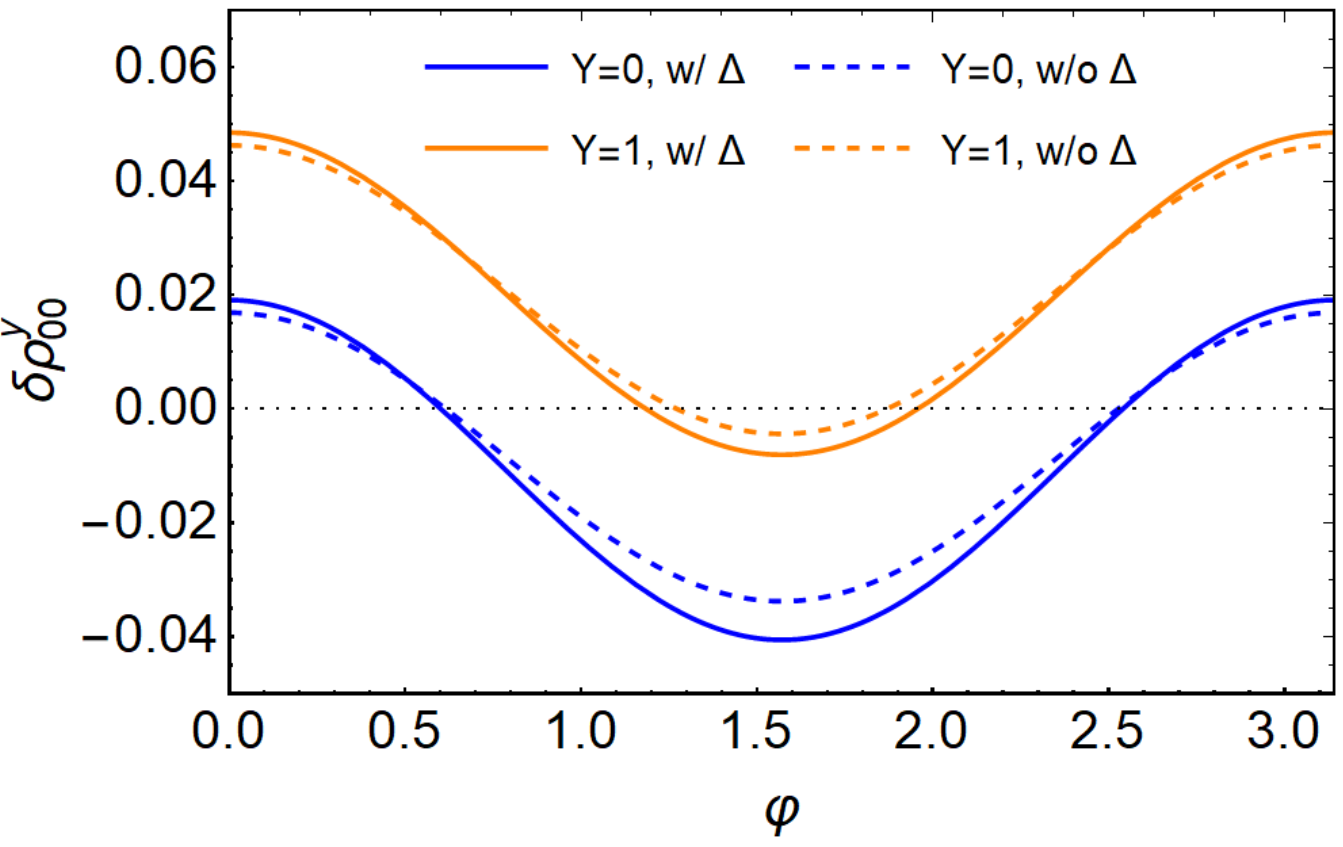}

\caption{\label{fig:Azimuthal_angle}Azimuthal angle $\varphi$ dependence
of $\delta\rho_{00}^{y}$ at rapidity $Y=0$ (blue lines) and $Y=1$
(orange lines) at 200 GeV. Solid lines are calculated using fluctuations
parameters $F^{2}$ and $\Delta$ given in the first paragraph of
Sec. \ref{sec:Numerical-results} and dashed lines are calculated
using the same $F^{2}$ but $\Delta=0$. }
\end{figure}

In order to study the transverse momentum dependence of $\rho_{00}^{y}$
in different rapidity regions, we present in Fig. \ref{fig:rho00_px_py}
the contour plot for the deviation $\delta\rho_{00}^{y}$ in $(p_{x},p_{y})$
plane at $Y=0$ (left panel) and $Y=1$ (right panel) at 200 GeV.
We observe a significant quadrupole structure: the mesons with $|p_{x}|\gg|p_{y}|$
have $\delta\rho_{00}^{y}>0$, while those with $|p_{x}|\ll|p_{y}|$
have $\delta\rho_{00}^{y}<0$. Such a structure is the result of Eq.
(\ref{eq:deviation_spin_alignemnt}) or (\ref{eq:deviation_pt_varphi_y}).

We also calculated the azimuthal angle dependence of $\delta\rho_{00}^{y}$
at a fixed transverse momentum $p_{T}=2$ GeV, see Fig. \ref{fig:Azimuthal_angle}.
For mesons at $Y=0$, $\delta\rho_{00}^{y}$ is positive at $\varphi=0$
and decreases to a negative minimum value at $\varphi=\pi/2$. The
curve shows a $\cos(2\varphi)$ behavior, as expected from Eq. (\ref{eq:deviation_pt_varphi_y}).
At a more forward rapidity $Y=1$, $\delta\rho_{00}^{y}$ is shifted
by a positive value relative to the $Y=0$ curve, which is also described
by Eq. (\ref{eq:deviation_pt_varphi_y}). The effects of the anisotropy
in strong force field fluctuations in the lab frame, which are quantified
by $\Delta$ in Eq. (\ref{eq:fluctuations_lab_frame}), are also shown
in Fig. \ref{fig:Azimuthal_angle}. We see that the effects of non-vanishing
$\Delta$ are small, implying that the azimuthal angle dependence
is dominated by the isotropic part $F^{2}$ in Eq. (\ref{eq:fluctuations_lab_frame}).


\section{Summary}

\label{sec:Summary}The momentum dependence of the $\phi$ meson's
global spin alignment in heavy-ion collisions is studied. According
to the relativistic quark coalescence model in the spin transport
theory for vector mesons \citep{Sheng:2022ffb,Sheng:2022wsy}, the
derivation from $1/3$ for $\rho_{00}$ for the $\phi$ meson is driven
by the anisotropy of local fluctuations of strong force fields in
the vector meson's rest frame, which can be related to fluctuations
in the lab frame through Lorentz transformation. From the geometry
of the quark-gluon plasma produced in heavy-ion collisions, it is
natural to assume that fluctuations are nearly isotropic in the lab
frame, with tiny anisotropy in the $z$-direction or beam direction.
By neglecting the anisotropy part, we derive an analytical expression
for $\rho_{00}^{y}-1/3$ which is proportional to $(p_{x}^{2}+p_{z}^{2})/2-p_{y}^{2}$,
indicating that the meson's motion along $x$- and $z$-directions
will enhance $\ensuremath{\rho_{00}^{y}}$, while the motion along
$y$-direction will decrease $\ensuremath{\rho_{00}^{y}}$. We then
predict the rapidity dependence of $\ensuremath{\rho_{00}^{y}}$ using
fluctuation parameters that are extracted from the STAR experiment
data on momentum-integrated $\ensuremath{\rho_{00}^{y}}$ \citep{STAR:2022fan}.
Our results show that $\ensuremath{\rho_{00}^{y}}$ has a negative
derivation from $1/3$ at mid-rapidity $Y=0$ and a positive derivation
at slightly forward rapidity $Y=1$. Predictions for the azimuthal
angle dependence of $\ensuremath{\rho_{00}^{y}}$ at these two rapidities
have also been made.

Although we assume that fluctuations in the lab frame contain an isotropic
part $F^{2}$ and an anisotropic part $\Delta$, we find in our calculation
that the contribution from the anisotropic part is negligible compared
with the isotropic part, as manifested in Fig. \ref{fig:Azimuthal_angle}.
This means that we may safely set $\Delta=0$ and still obtain nearly
the same results as Fig. \ref{fig:spin_alignment_rapidity}. The momentum
dependence is mainly caused by the broken rotational symmetry due
to the motion of the $\phi$ meson relative to the bulk matter. Such
a mechanism can be tested in future experiments.


\section*{Acknowledgement}

We thank Jinhui Chen, Aihong Tang, Gavin Wilks and Xu Sun for helpful
discussions. This work is supported in part by the National Natural
Science Foundation of China (NSFC) under Grant No. 12135011 and 12075235,
by the Strategic Priority Research Program of the Chinese Academy
of Sciences (CAS) under Grant No. XDB34030102.

\bibliographystyle{apsrev}
\bibliography{rapidity-phi-spin-3}

\end{document}